\documentclass[%
 reprint,
 amsmath,amssymb,
 aps,
]{revtex4-2}
\usepackage{graphicx}% Include figure files
\usepackage{dcolumn}% Align table columns on decimal point
\usepackage{bm}% bold math
\usepackage[colorlinks=true, citecolor=blue, urlcolor=blue, linkcolor=magenta]{hyperref}
\usepackage{float}
\usepackage[bottom]{footmisc}
\usepackage{booktabs}
\usepackage{subfigure}
\usepackage{caption}
\usepackage{xcolor}
\usepackage{amssymb}
\usepackage{amsmath}
\usepackage{url}
\usepackage{natbib}

\newcommand{\physrep}{Phys. Rep.}

\newcommand{\mnras}{Mon. Not. Roy. Astron. Soc.}

\newcommand{\aap}{Astron. Astrophys.}

\begin{document}

\title{Hadronic Origin of Sub‑PeV Gamma‑Ray Emission from LHAASO J0621+3755}

\author{Sonali Sahoo}
% \altaffiliation[]%Lines break automatically or can be forced with \\\email{m22ph003@iitj.ac.in}
 \email{p23ph0014@iitj.ac.in}
\author{Ankan Roy}%
 \email{m22ph003@iitj.ac.in}
\author{Kritika Yadav}
 \email{m22ph201@iitj.ac.in}
\author{Reetanjali Moharana}
 \email{reetanjali@iitj.ac.in}
\affiliation{
 Department of Physics, Indian Institute of Technology Jodhpur\\
 NH-62, Karwar, Rajasthan-342037
}

%\collaboration{}%\noaffiliation


\date{\today}% It is always \today, today,

\begin{abstract}
Very High Energy (VHE) gamma-rays in pulsars, and their surrounding halos, are interpreted to originate from the leptonic channel, electromagnetic interactions through electron inverse Compton (IC) scattering. In the hadronic scenario, TeV-PeV gamma-rays are generated from the decay of neutral pions, which are produced from cosmic rays(CR) protons interacting with the ambient medium. Recent observations of sub-PeV gamma-rays from the halo of the pulsar PSR J0622+3749 by the Large High Altitude Air Shower Observatory Kilometer-Square Array (LHAASO-KM2A) provide an opportunity to investigate the underlying emission mechanisms. Previous studies have shown that the observed emission can be consistently explained within a leptonic framework by the slow diffusion of electrons. In this work, we explore an alternative explanation based on the hadronic scenario through the proton-proton ($pp$) interaction channel, incorporating the observation of VHE gamma-rays at 7 TeV by the High-Altitude Water Cherenkov detector (HAWC). To model the observed gamma-ray spectrum, ranging from $\sim 7~\mathrm{TeV}$ up to $200~\mathrm{TeV}$, the required CR proton luminosity is found to be $\eta_p \sim 0.14$ of the spin-down luminosity of PSR J0622+3749. This scenario assumes that protons propagate in a one-zone superdiffusive environment, characterized by a diffusion index $\alpha = 1.05$, within an ambient of density, $1~\mathrm{cm}^{-3}$.

\end{abstract}

\maketitle

%\tableofcontents

\section*{\label{intro}Introduction} %\centering

GeV-TeV gamma-rays have been observed from several pulsar wind nebulae (PWNs), such as Vela \cite{HESS:2019jxy}, MSH 15-52 \cite{ HESS:2005lqd}, and Crab \cite{2020NatAs...4..167H} using Imaging Atmospheric Cherenkov Telescopes (IACTs). These gamma-rays are generally interpreted within a leptonic framework. The rotational energy of the pulsar in the PWN drives an ultrarelativistic magnetized wind containing electron-positron pairs \cite{pacini1973evolution,10.1093/mnras/167.1.1} and protons (or more generally ions) from the surface of the rapidly spinning pulsar \cite{Arons_Tavani_1994}, into the ambient medium. A termination shock is formed by the wind in the medium, where the charged particles accelerate \cite{Gaensler:2006ua}. The accelerated $e^{-}/e^{+}$, diffuse into the surrounding halo and produce gamma-rays through electromagnetic (EM) processes. In this picture, synchrotron radiation accounts for the emission from radio to X-ray wavelengths, while GeV-TeV gamma-rays originate from IC scattering of ambient photons by the lepton pairs \cite{HESS:2005lqd, PhysRevLett.15.577,1969ApJ...155..429R,10.1093/mnras/291.1.162}.

Recent observations have extended the detected gamma-ray spectrum to sub-PeV energies in several astrophysical environments, including PWN and their surrounding halo. In particular, PSR J0622+3749/LHAASO J0621+3755, located at a pseudo-distance $d_L = 1.6~\mathrm{kpc}$~\cite{parkinson2010eight}, has been detected by LHAASO-KM2A above $10~\mathrm{TeV}$ up to energies exceeding $100~\mathrm{TeV}$ as an extended source~\cite{PhysRevLett.126.241103}. The best-fit position of LHAASO J0621+3755 is $\mathrm{RA} = 95.47^\circ$, $\mathrm{Dec} = 37.92^\circ$, with an angular extension of $0.40^\circ \pm 0.07^\circ$. The centroid of the VHE emission is offset by $0.11^\circ \pm 0.12^\circ$ from the middle-aged pulsar PSR J0622+3749 above $25~\mathrm{TeV}$, suggesting a physical association. Further, a search for GeV emission using Fermi-Large Area Telescope (LAT) in the energy range $15$-$500~\mathrm{GeV}$ yields no significant detection, providing only upper limits. 

Subsequent multiwavelength observations with XMM-Newton in March and April 2023, with a total exposure of $74~\mathrm{ks}$, resulted in a non-detection in the $2$-$7~\mathrm{keV}$ band, providing an upper limit on the X-ray flux for the sources. Similarly, Very Energetic Radiation Imaging Telescope Array System (VERITAS) observations conducted over $\sim 40$ hours during 2022 and 2023, covering the energy range $0.3$-$10~\mathrm{TeV}$, reported no significant detection and placed upper limits in the energy range.

The absence of X-ray emission, when interpreted within a leptonic synchrotron scenario, constrains the magnetic field in the halo region to $\lesssim 1~\mu\mathrm{G}$ ~\cite{VERITAS:2025xjd}. The VHE emissions are modeled in reference \cite{PhysRevLett.126.241103, Fang:2021qon,VERITAS:2025xjd} with the leptonic channel, where the relativistic $e^+$/$e^-$ in the halo of PSR J0622+3749 upscatter ambient photon fields. The modeled target photon fields are three graybody components with temperatures of $5000^\circ\,\mathrm{K}$, $20^\circ\,\mathrm{K}$, and $2.73^\circ\,\mathrm{K}$, and corresponding energy densities of $0.3\,\mathrm{eV\,cm^{-3}}$ and $0.26\,\mathrm{eV\,cm^{-3}}$. The modeling also requires a slow diffusion of relativistic electrons with a diffusion coefficient of $8.9 \times 10^{27}\,\mathrm{cm^2\,s^{-1}}$.

The nearest TeV counterpart to LHAASO J0621+3755 is the source 3HWC J0621+382, reported in the third HAWC catalog~\cite{HAWC:2020hrt}. This source lies at an angular separation of $0.31^\circ \pm 0.32^\circ$ from LHAASO J0621+3755 and $0.42^\circ$ from PSR J0622+3749. The differential flux of 3HWC J0621+382 is derived assuming a disk-like morphology with a radius of $0.5^\circ$, using 1523 days of HAWC observations. The spectrum is well described by a power-law with index $-2.41^{+0.12}_{-0.13}{}^{, \, +0.08}_{, \, -0.01}$ (statistical and systematic uncertainties) and a normalization of $8.9^{+1.4}_{-1.5}{}^{, \, +2.6}_{, \, -1.2} \times 10^{-15}~\mathrm{TeV^{-1}\,cm^{-2}\,s^{-1}}$ at a pivot energy of $7~\mathrm{TeV}$.

Reference~\cite{PhysRevLett.126.241103} suggests the source 3HWC J0621+382 is associated with the blazar candidate of uncertain type (bcu) 4FGL J0620.3+3804/GB6 J0620+3806, with an angular separation of $0.22^\circ$, rather than with the halo of PSR J0622+3749. Hence, while modeling the observations of 3HWC J0621+382 has not been taken into account. Additionally, the modeling of electron diffusion predicts a significantly higher flux in the $1$-$10~\mathrm{TeV}$ range than that observed for 3HWC J0621+382. 

To further investigate, we perform a dedicated Fermi-LAT analysis (section~\ref{text-SED-3804}) and present the spectral energy distribution (SED) of 4FGL J0620.3+3804 in figure~\ref{SED-3804}. The spectrum is well described by a power-law, $dN/dE = N_0 \left(E/E_0\right)^{\gamma}$, with $\gamma = -2.54$, $N_0 = 6.99 \times 10^{-14}~\mathrm{MeV^{-1}\,cm^{-2}\,s^{-1}}$, and $E_0 = 1809.2~\mathrm{MeV}$. The resulting spectrum exhibits minimal emission in the TeV energy range.

Notably, the observed spectrum of LHAASO J0621+3755 exhibits a rise in flux between $\sim 7$ and $10~\mathrm{TeV}$, followed by a power law behavior extending up to $\sim 200~\mathrm{TeV}$. In light of the above discussion, if 3HWC J0621+382 is physically associated with the halo of LHAASO J0621+3755, an alternative mechanism is required to explain the gamma-ray emission over the full energy range from $\sim 7~\mathrm{TeV}$ to sub-PeV energies.

 Hadronic processes provide an alternative channel for producing sub-PeV gamma-rays in PWN environments. Previous studies indicate that, during the early stages of pulsar evolution, the injected particle composition includes ions, whose abundance remains an open question. A prediction by ~\cite{Fang:2013cba, Kotera:2015pya} suggests the outflow wind consists of $\sim 50\%$ protons, $\sim 30\%$ CNO nuclei, and $\sim 20\%$ Fe. Like electrons, these ions can also accelerate in the termination shock. However, due to slow diffusion and no appreciable energy losses, unlike leptons, protons accumulate in the PWN.  

 Subsequently, high-energy gamma-rays originate from the decay of neutral pions produced in interactions of accelerated protons with ambient photons ($p\gamma$) or matter ($pp$). Both $p\gamma$ and $pp$ channels are expected to generate comparable energy fluxes in gamma-rays and their neutrino counterparts. However, for old or middle-aged PWNs (note, PSR J0622+3749 has an estimated age of $t_{\rm age} \approx 207.8\,\mathrm{kyr}$), the ambient photon fields are typically characterized by temperatures below $5000^{\circ}$ K~\cite{Martin:2022blv,2009RAA.....9..449Q, Torres:2013jha}, which suppresses the efficiency of the $p\gamma$ channel in producing TeV gamma-rays in their halos. Additionally, the proton spectrum can be hard, leading to higher-energy gamma-ray production than in the leptonic scenario. Several studies have considered purely hadronic scenarios, particularly involving the $pp$ channel, to explain gamma-ray emission from PWNs. Notable examples include Vela X~\cite{Horns:2006ku}, SNR G24.7+0.6/PWN HESS J1837-069~\cite{Banik:2021dvz}, and HESS J1857+026~\cite{2012A&A...544A...3R}.

A more comprehensive description of PWNs can be achieved within a lepto-hadronic framework, where multiwavelength emission from radio to X-rays is attributed to synchrotron radiation of leptons, and GeV-TeV gamma-rays arise from IC scattering. In contrast, sub-PeV gamma-rays are explained by hadronic interactions; for example, see \cite{Bednarek:2003tk, Arons:1998pg}. 
Motivated by these features, we propose that the sub-PeV gamma-rays from LHAASO J0621+3755 originate from a channel distinct from the leptonic scenarios previously invoked across GeV–PeV energies.

The paper is organized as follows. In section~\ref{fermi-lat}, we present the Fermi-LAT analysis of the region around PSR J0622+3749. The spectral energy distribution (SED) of PSR J0622+3749/4FGL J0622+3749, modeled as a point source, is discussed in section~\ref{SED-3749}. An extended source analysis, motivated by the morphology of LHAASO J0621+3755, is presented in section~\ref{fermi-lat analysis of 3755}. To improve source localization, we perform a sky map analysis in section~\ref{skymap}. The spectral analysis of the bcu 4FGL J0620.3+3804 is described in section~\ref{text-SED-3804}.

In section~\ref{pp model}, we outline the $pp$ interaction model and the corresponding gamma-ray flux calculation. We also describe the diffusion framework adopted in this work. The particle transport equation is solved using the GAMERA (Gamma-ray and Astroparticle Modeling for Emission and Radiation Analysis) framework~\cite{Hahn:2016CO}. In section~\ref{Analysis}, we present the results and discuss the implications of the model, including estimates of the associated neutrino flux and the expected number of neutrino events, providing a testable prediction for the proposed hadronic scenario.

\section{Fermi-LAT data analysis} \label{fermi-lat}
The Fermi-LAT is a highly sensitive $\gamma$-ray telescope covering energies from $\sim 100$ MeV to $\sim 300$ GeV, with a 2.4 sr field of view that enables full-sky coverage every 3 hours. In this section, we provide the Fermi-LAT data analysis of the region around PSR J0622+3749, including the point source PSR J0622+3749, the extended source LHAASO J0621+3755, and the blazar 4FGL J0620.3+3804/GB6 J0620+3806, to explore the MeV-GeV emissions.

For our analysis, we used the LAT data for a search radius of $15^\circ$ centered at PSR J0622+3749,  covering the time period from  August 15, 2008 to February 1, 2020, approximately 12 years, with the energy range 300 MeV to 1 TeV (note that the Fermi-LAT data above 300 GeV may have less sensitivity, however, to check spectral form, we have extended upto 1 TeV) from the Pass 8 \texttt{SOURCE} event class, represented by the \texttt{P8R3\_SOURCE\_V3} instrument response function. To minimize contamination from Earth limb \(\gamma\)-rays, we exclude events with zenith angles greater than \(90^\circ\). 

\subsection{\label{SED-3749} SED for PSR J0622+3749/4FGL J0622+3749}
We conducted SED analysis for the point source PSR J0622+3749 to study its energetics. This analysis was performed using Fermipy version 1.2.2 \cite{fermipy}, with the source located at RA = $95.54^\circ$ and Dec = $37.82^\circ$. We generated the SED using binned likelihood analysis, based on photon data and the spacecraft file.

\begin{figure}[t]
   \centering
   \includegraphics[width=0.98\linewidth]{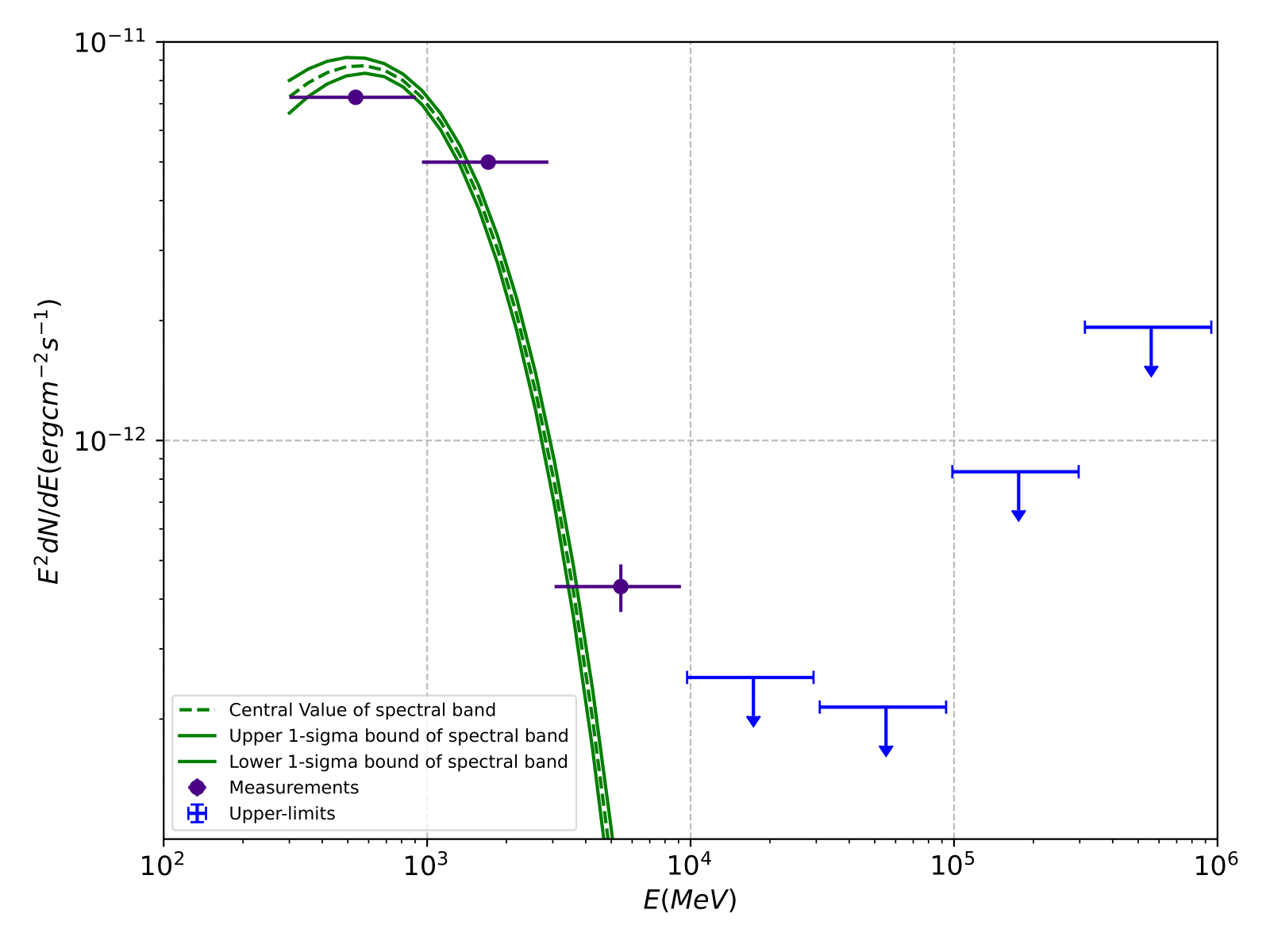}
   \caption{SED of pulsar PSR J0622+3749/4FGL J0622.2+3749  with Fermi-LAT as a point source.}
   \label{SED}
\end{figure}

In the binning section, we chose a region of interest (RoI) of width \(10^\circ\) for a bin size of \(0.1^\circ \times 0.1^\circ\) in instrument coordinates\footnote{\url{https://fermipy.readthedocs.io/en/latest/config.html}}. We then selected \(\text{evclass}=128\) and \(\text{evtype}=3\), which means that we are taking both photon conversion types FRONT and BACK, with different angular resolution and effective area for the corresponding Instrument Response Function (IRFs)
\footnote{\url{https://fermi.gsfc.nasa.gov/ssc/data/analysis/documentation/Cicerone/Cicerone_LAT_IRFs/IRF_overview.html}}. We have selected the Relational filter expression (filter) value to be \texttt{(DATA\_QUAL > 90 \&\& LAT\_CONFIG == 1)}. The diffuse models used are \texttt{gll\_iem\_v07.fits} and \texttt{iso\_P8R3\_SOURCE\_V3\_v1.txt} \footnote{\url{https://fermi.gsfc.nasa.gov/ssc/data/access/lat/BackgroundModels.html}}.
We also used fermipy catalog file \texttt{4FGL-DR3} so that it can automatically generate a model for our region using catalog location and spectral information. We removed sources with test statistics \texttt{TS < 3}. In the next step, we free the normalization factor for all sources within $5^\circ$ of the RoI center. We use free sources with \texttt{TS > 10} along with the isotropic and galactic diffuse components. After examining the results, we fixed the parameter and generated SED for 4FGL J0622.2+3749 shown in figure \ref{SED} with the three highest energy bins \texttt{TS < 20}, while the flux upper-limits \texttt{TS < 20}. The model fitted to the SED follows PLSuperExpCutoff with a power law index of $2.31 \pm 0.053$ and a peak at $710.63\pm 41.16$ MeV. Hence, the pulsar is unlikely to produce non-thermal TeV photons. Modelings in \cite{2005PhRvL..94r1101L,2009MNRAS.395.1371B} suggest the possibility of pulsars producing TeV gamma and neutrino emissions. The spectral fitting here constrains these models for the PSR J0622+3749.

\subsection{\label{fermi-lat analysis of 3755}Fermi-LAT Data Analysis of LHAASO J0621+3755}

To analyse the extended emission from PSR J0622+3749, we chose an RoI \(10^\circ\) and used a bin size of \(0.1^\circ \times 0.1^\circ\) in instrument coordinates. Following \cite{PhysRevLett.126.241103}, photons in the energy range 15–500 GeV were selected, and a binned likelihood analysis was performed using two logarithmically spaced energy bins per decade. The energy bins with \texttt{TS < 10} are shown as upper limits in figure ~\ref{Wholesed}.

The source size and position could differ at GeV energies, so we conducted at least 50 trials to determine the position with the highest TS value. We tested a set of grid positions within RA (\(90.75^\circ - 100.25^\circ\)) and Dec (\(35^\circ - 42^\circ\)) with a width around \(0.4^\circ\) using \texttt{RadialGaussian} spatial model and \texttt{PlSuperExpCutoff}(SupExp) / \texttt{Power Law}(PL) as spectrum type to obtain a comparatively high TS for the source. Despite these trials, the maximum TS value we have is always less than 8. The upper limits are derived with \texttt{TS\_threshold = 10} and are shown in the figure \ref{Wholesed}. As mentioned above, the halo size does not exceed 1°. 
 Since the study is limited to a source region of only $0.4^{\circ}$, one can assume that the Fermi-LAT background model is not absorbed. Hence, the significance of this analysis study may result in suppression of the signal above the GeV energy band for the LHAASO J0621+3755 source. 
\begin{figure}[t]
   \centering
   \includegraphics[width=0.98\linewidth]{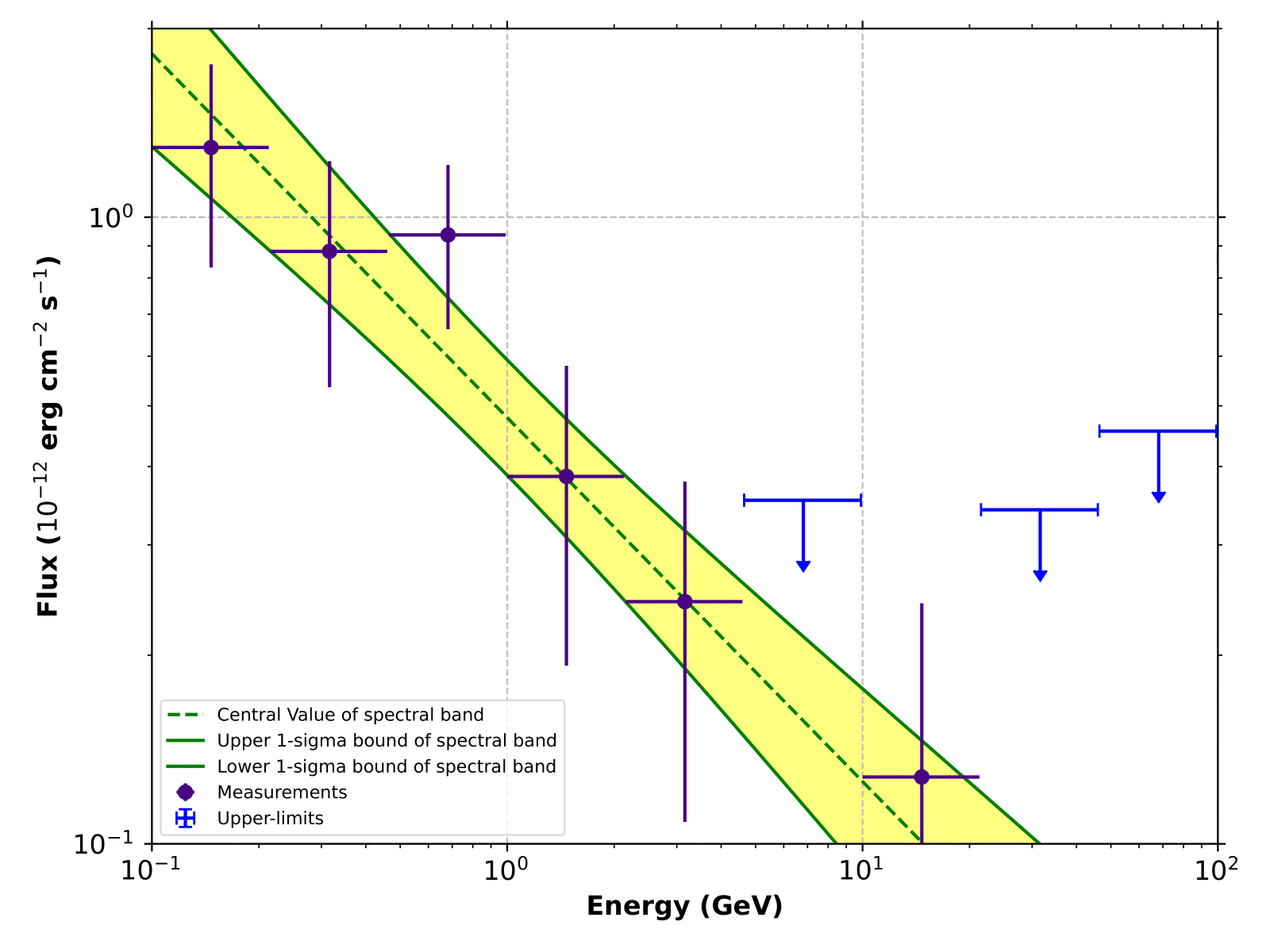}
   \caption{SED of bcu 4FGL~J0620.3+3804 with Fermi-LAT; the uncertainty is represented by a butterfly-shaped shaded region.}

   \label{SED-3804}
\end{figure}
\subsection{\label{skymap}Sky Map}
To investigate the location of the events in detail, we performed a sky map analysis around the pulsar PSR J0622+3749, considering a RoI of approximately $1.5^{\circ}$. We have selected \texttt{evclass=128}, \texttt{evtype=3}, and \texttt{CCUBE} (counts cube) option from \texttt{gtbin} tool of Fermi Science Tools version 2.2.0 to generate sky maps on different energy scales. The analysis utilized data collected between January 1, 2011, and February 4, 2024, over the same energy range as adopted previously, with a search radius of $15^\circ$ centered on PSR J0621+3749.

\begin{figure*}[t]
    \centering
    \begin{minipage}[c]{0.48\linewidth}
        \centering
        \includegraphics[width=1.0\linewidth]{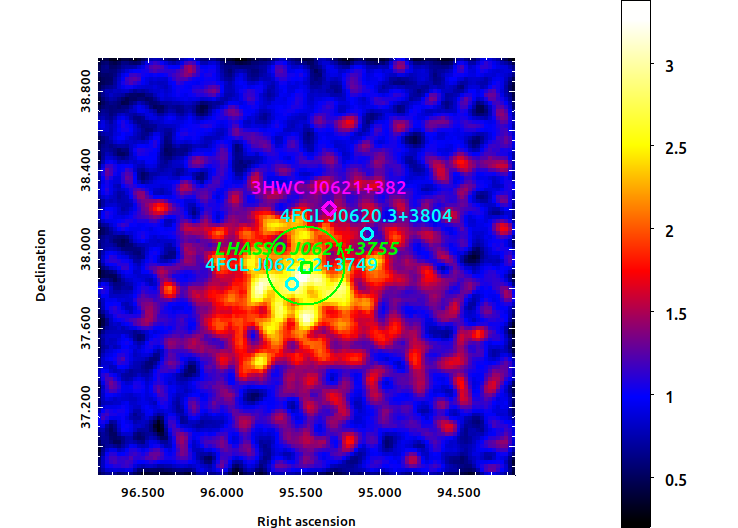} 
    \end{minipage}
    \hspace{0.01\linewidth}
    \begin{minipage}[c]{0.48\linewidth}
        \centering
        \includegraphics[width=1.0\linewidth]{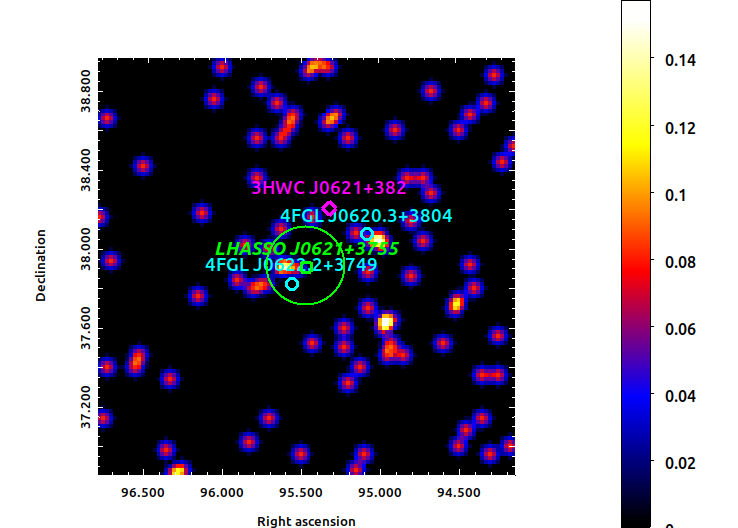}
    \end{minipage}
    \caption{\raggedright\setlength{\rightskip}{0pt}\setlength{\leftskip}{0pt}Count map in units of counts per pixel for the region around RA = \(95.47^\circ\) Dec = \(37.92^\circ\) for two logarithmically equi-spaced energy bins between 300 MeV to 300 GeV (300 MeV-9.487 GeV and 9.487 GeV-300 GeV, respectively, left and right). The magenta diamond marks the location of the 3HWC J0621+382. The green square and circle denote the best fit and 1 $\sigma$ range of the location of the LHAASO J0621+3755 source. The cyan circle shows the locations of two 4FGL sources.}
    \label{fig-1}
    \label{fig:sky_maps}
\end{figure*}

Figure \ref{fig-1} presents the Aitoff projection skymap of $3^{\circ} \times 3^{\circ}$ around PSR J0622+3749 at RA= $95.47^\circ$ and Dec = $37.92^\circ$, with a pixel size of $0.02^{\circ}$ in Celestial coordinates (CEL). The plots on the left and right of figure \ref{fig-1} show the energy sky map, with two logarithmically equispaced energy bins spanning 300 MeV to 300 GeV. The positions of known sources are also indicated. The map clearly shows that the halo region exhibits negligible emission at GeV energies.

%\begin{table}[h!]
%\centering
%\caption{Source Details for which 95\% upper limits are calculated}
%\begin{tabular}{ll}
%\toprule
%\text{RA/DEC}           & 94.985 / 37.907 \\
%\text{GLON/GLAT}        & 175.600 / 10.600 \\
%\text{TS}               & 0.69 \\
%\text{Npred}            & 2.46 \\
%\text{Flux}             & $4.716 \times 10^{-12} \pm 6.41 \times 10^{-12}$ \\
%\text{EnergyFlux}       & $4.084 \times 10^{-7} \pm 5.55 \times 10^{-7}$ \\
%\text{SpatialModel}     & \texttt{RadialGaussian} \\
%\text{SpectrumType}     & \texttt{PLSuperExpCutoff} \\

%\text{Spectral Parameters} & \\
%\quad \text{Prefactor}  & $1.105 \times 10^{-14} \pm 1.5 \times 10^{-14}$ \\
%\quad \text{Index1}     & -1.5  \\
%\quad \text{Scale}      & 1000  \\
%\quad \text{Cutoff}     & $1.5 \times 10^{14}$ \\
%\quad \text{Index2}     & 1 \\
%\bottomrule
%\end{tabular}
%\end{table}

\subsection{\label{text-SED-3804}SED analysis of 4FGL J0620.3+3804}
The bcu source 4FGL J0620.3+3804 is proposed to be associated with the 3HWC J0621+382. Here, we analyzed Fermi-LAT data for the source over approximately 11 years, from January 1, 2011 to January 1, 2022, with energies ranging from 300 MeV to 1 TeV. We performed a binned likelihood analysis of the events with a bin size of \(0.1^\circ \times 0.1^\circ\). RoI is centered at RA = 
$95.08^\circ$ and Dec = $38.08^\circ$ (see \footnote{4FGL-DR4 positions and source names: \url{https://fermi.gsfc.nasa.gov/ssc/data/access/lat/14yr_catalog/gll_psc_v32.reg}}), with a radius of \(5^\circ\).
Using a similar analysis (label \ref{SED-3749}), we generated the SED for 4FGL J0620.3+3804 and presented it in figure \ref{SED-3804}. The best-fit SED model follows a power law with a spectral index of -2.54. At an energy of 10 GeV the flux becomes, $10^{-13} \text {erg} \, \text {cm}^{-2} \, \text{s}^{-1}$, whereas the 3HWC J0621+382 has a value nearly $7 \times 10^{-13}\text {erg} \, \text {cm}^{-2} \, \text{s}^{-1}$ at 7 TeV. Hence, the  3HWC J0621+382 will be difficult to explain by the bcu GB6 J0620+3806/4FGL J0620.3+3804.

\section{\label{pp model}Modeling of the VHE emission from PSR J0622+3749}

The studies cited in \textcolor{blue}{\cite{PhysRevLett.126.241103, Fang:2021qon, PhysRevD.104.123017}} interpret LHAASO J0621+3755 as originating from IC scattering of the slow diffusion zone trapped relativistic electrons around PSR J0622+3749. The modeling exhibits both a one-zone superdiffusion model and a two-zone normal diffusion model to explain the sub-PeV $\gamma$-rays. Further around 
the X-ray band,  the XMM-Newton upper limit constrains the magnetic field, B, of the halo to $\leq 1\,\mu\mathrm{G}$ modeled from synchrotron emission. However, in all the above modeling, the 3HWC J0621+382 has been excluded, assuming that the HAWC observed TeV $\gamma$-rays could be generated by a different source, bcu-4FGL J0620.3+3804. But our study of the bcu energetics, in section \ref{text-SED-3804}, suggests the source has a negligible chance of producing TeV $\gamma$-rays as observed by HAWC. A modified leptonic origin may simultaneously explain both 3HWC J0621+382 and LHAASO J0621+3755, for example, modeling of PWN G359.95-0.04 \cite{Hinton:2006zk}. In this lepton model, a high minimum energy for the injected electrons is required, allowing the emission to display a dip at lower gamma-ray energies. It is important to note that the necessity of such a high minimum energy is subject to debate.

Here, we propose a hadronic production mechanism to explain the VHE emissions in the energy range of $\sim$ 7 TeV, extending to 200 TeV. In particular, we modeled the interaction of termination shock-accelerated protons \cite{Amato:2006ts} with the ambient medium, producing neutral pions that eventually decay into VHE photons. This model can dominate the lepton channel at higher energies, particularly beyond several TeV. This is primarily due to the hard spectra of protons and the more rapid cooling of high-energy leptons. Compared to leptons, protons have a longer cooling time and continue to accumulate in the surrounding environment after being accelerated. 

Protons lose energy in such an environment primarily through $pp$ interactions and synchrotron radiation. The average proton cooling time scale is becomes,  $\frac{1}{t_{cool}} = \frac{1}{t_{pp}} + \frac {1}{t_{sync}}$. The $pp$ channel time scale, $t_{pp} =1/(n_{H} \sigma_{pp}c)$, for an average ambient density, $\langle n_{H} \rangle=1~\text{cm}^{-3}$ and inelastic scattering cross-section $\sigma_{pp}$ that depends on $E_{c} = \left(\frac{E_{\pi}}{K_{\pi}}\right) + m_{p}c^{2}$,$ 
E_{\pi,\text{th}} = E_{\gamma} + \frac{m_{\pi}^{2}c^{4}}{4E_{\gamma}}$, result $\sim 22000$ kyr at $E_p$ = 100 TeV, as explained in  \cite{Kelner:2006tc}. The proton synchrotron cooling time scale in a magnetic field, B, is $t_{p,sync}=1.5 \times 10^{15} \, \text{kyr}\, B_{\mu G}^{-2} E_{p,100\text{TeV}}$ \cite{Aharonian:2000pv}. These calculations show a long cooling time for protons, confirming their survival after being extracted from the surfaces of neutron stars. Furthermore, the different spectral forms of protons and electrons result in the generation of photons with energies exceeding 10 TeV \cite{Schroer:2023aoh}. Several examples of PWN modeling using $pp$ have claimed the production of sub-PeV photons, citing the same reasons as above, for example, as in \cite{Horns:2006ku}.

The $pp$ interaction channel leads to the creation of secondary neutral pions ($\pi^0$) and charged pions ($\pi^{\pm}$), which subsequently decay into gamma-rays ($\gamma$-rays) and neutrinos ($\nu_{e,\mu}$), respectively. The efficiency of this mechanism depends on the target proton density in the ambient medium and the energy distribution of the accelerated particles. 

In this work, we have taken the time-dependent injected proton spectra, with energy spectral index $\alpha_p$, as adopted for electro-injection spectra in \cite{Fang:2021qon}. 
\begin{equation}
Q_{p}(N_p,\gamma_p,t) = N_p\left[\frac{t_{sd}+t_{age}}{t+t_{sd}}\right]^{2},
\end{equation}
considering,
\begin{equation}
N_p(\gamma_p)=N_{0} \gamma_p^{-\alpha_p}\exp\!\left[-\left(\frac{\gamma_p}{\gamma_{\mathrm{cut}}}\right)\right],
\end{equation}
where $N_{0}$ is the normalisation factor (in units of $\mathrm{erg}^{-1}\,\mathrm{s}^{-1}$). A Lorentz boost, $\gamma_p = \frac{E_p}{m_p}$, where $E_p$ and $m_p$ are the relativistic energy and rest mass of the proton; additionally, $\gamma_{cut}$ corresponds to $E_p = 250$ TeV.  Here, the spectra are considered to be proportional to the spin-down luminosity, which is $\propto (1+t/t_{sd})^{-2}$, for a spin-down time ($t_{sd}$) of 10 kyr \textcolor{blue}{\cite{PhysRevLett.126.241103}}. The transport equation of the proton in the halo is,
\begin{equation}
\frac{\partial N_p(\gamma_p)}{\partial t}
=
Q_p(N_p,\gamma_p,t)
-
\frac{\partial}{\partial \gamma_p}\left( b N_p(\gamma_p) \right)
-
\frac{N_p(\gamma_p)}{t_{\rm esc}(E_p,\alpha)}
\label{eq:transport}.
\end{equation}

Here, $b \equiv -\mathrm{d}E_p/\mathrm{d}t$ represents the energy loss rate. One can assume b = 0 because of the extended cooling time of protons. Additionally, the propagation of the injected proton spectrum into the halo is governed by diffusion. We have considered here a time-dependent one-zone Kolmogorov diffusion scenario. The time-dependent diffusion generally follows a mean squared displacement (MSD) of particles as $\langle r^{2}\rangle \propto t^{\alpha }$ \cite{2000PhR...339....1M,2013ApJ...778...35Z}. For $1< \alpha <2 $, it becomes superdiffusion, whereas the normal diffusion follows $\alpha =1 $. Since the PSR J0622+3749 is in its middle age, the Kolmogorov scenario, where the energy-dependent diffusion, $D(E_p)\propto E_p^{1/3}$, has the least contribution. In summary, the escape time of protons in the PWN environment considered in our scenario follows \cite{2000PhR...339....1M}, 
\begin{equation}
t_{esc}(E_p,\alpha)= \left[\frac{R^2}{2D_{\alpha}(E_p)}\right]^{1/\alpha}
\end{equation}

R is the halo size and is considered to be 50 pc for our modeling. $D_\alpha(E_p)=D_0\left(\frac{E_p}{160 \, \text{TeV}}\right)^{1/3}$. Here, we have considered $\alpha$ = 1.05. This is equivalent to the values obtained for the electron propagation in the IC-emitted model for the same source \cite{PhysRevLett.126.241103}. Although the proton is much more massive than the electron, we can still assume that the diffusion parameters are the same, but only in a relativistic scenario.

\begin{figure}[t]
   \centering
   \includegraphics[width=0.895\linewidth]{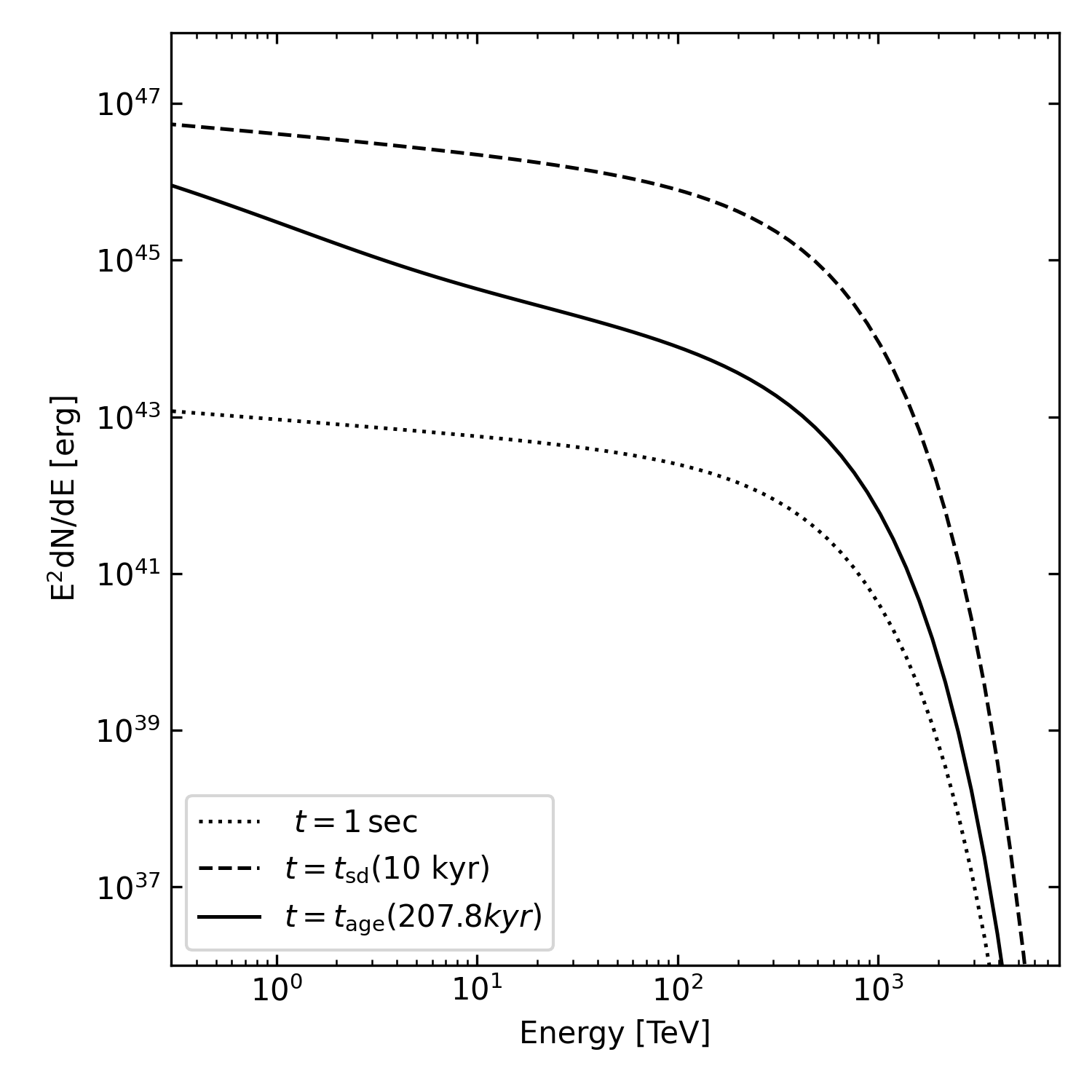}
   \caption{\raggedright\setlength{\rightskip}{0pt}\setlength{\leftskip}{0pt}Time evolution of the proton energy spectrum in the PWN J0622+3749. The dotted, dashed, and solid curves correspond to $t = 1\,\mathrm{s}$, $10\,\mathrm{kyr}$,  and $207.8\,\mathrm{kyr}$, respectively.
}
   \label{time}
\end{figure}

We solve this equation to get the survived proton spectra using GAMERA \cite{Hahn:2016CO} within the region of R. figure \ref{time} shows the injection proton spectra at time 1 sec, with a dotted line, and at the spin-down time with a dashed line, compared to the surviving protons till the age of the PWN J0622+3749 with a solid line. This suggests the accumulation of protons in the halo over time. 

We again used GAMERA to calculate the VHE gamma-ray emission from the halo at $d_{L}$, through the $pp$ channel using PYTHIA 8.18 \cite{Sjostrand:2007gs}. 
A comparison with the analytical form of $\gamma$-ray photon flux resulting from the \textit{pp} interaction via $\pi^0$ decay can also be attempted, following \cite{1996A&A...309..917A,Moharana_2016},
\begin{equation}
\frac{dN_{\gamma}}{dE_{\gamma}} = 
\frac{2 c \tilde{n} \langle n_{H}\rangle }{4\pi D_{L}^{2} K_{\pi}}
\int_{E_{\pi,\text{th}}}^{\infty} 
\frac{dE_{\pi}}{\sqrt{E_{\pi}^{2} - m_{\pi}^{2}}}
\, \sigma_{pp}(E_{c}) \, N_{p}(E_{c}).
\end{equation}
Here, $\tilde{n}$ represents the number of pions produced per interaction, which is assumed to be one. The factor $K_{\pi}$, representing pion multiplicity, is generally taken to be 0.17.

\begin{figure}[h]
\centering
\includegraphics[width=0.986\linewidth]{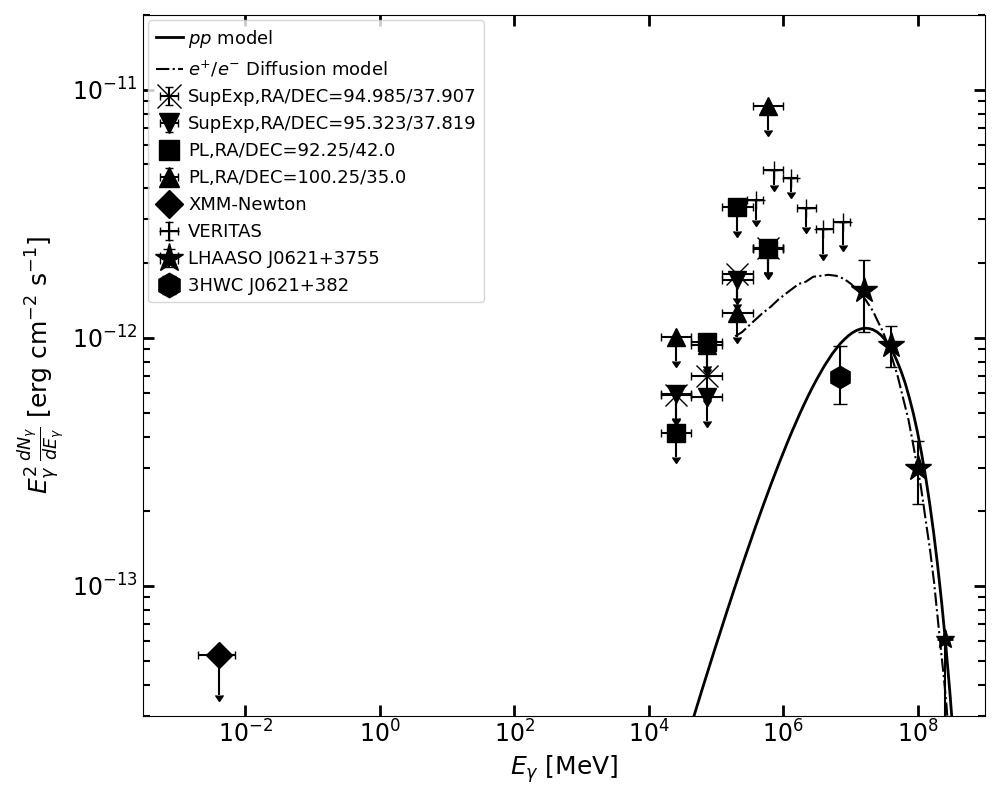}
\caption{\raggedright\setlength{\rightskip}{0pt}\setlength{\leftskip}{0pt} The Fermi-LAT upper limits for the PL and SupExp models, LHAASO J0621+3755, are presented within the energy range of $15\,\mathrm{GeV}$\ to\ $500\,\mathrm{GeV}$. The X-ray (XMM-Newton) and gamma-ray (VERITAS) upper limits are taken from \cite{VERITAS:2025xjd}. The solid line represents the modeling for the ($pp$) interactions, while the \(e^{\pm}\) diffusion model is shown with a dash-dotted line, as referenced in \cite{PhysRevLett.126.241103}.}
\label{Wholesed}
\end{figure}

The parameters  $\alpha_p$ and $N_{0}$ are optimised by modeling the observed VHE $\gamma$-rays by HAWC and LHAASO and considering the upper limits of Fermi-LAT and the VERITAS observatory as cited in \cite{VERITAS:2025xjd}. Figure \ref{Wholesed} illustrates the hadronic model proposed in this study, represented by a solid line. Star points indicate the photon counts observed by LHAASO-KM2A, while circular points represent the observations for 3HAWC J0621+382 \cite{HAWC:2020hrt}, along with the systematic and statistical uncertainties obtained through error propagation. Additionally, the SED displays the upper limits recorded by Fermi-LAT, as detailed in subsection \ref{fermi-lat analysis of 3755}, as well as the electron diffusion model suggested by the LHAASO collaboration \cite{PhysRevLett.126.241103}. The parameters used to derive our model are listed in table \ref{tab:parameters_diffusion}.

Using the model parameters, we calculated the luminosity of the protons, expressed as:

\[
L_{p} = \int_{\gamma_{p,\text{min}}}^{\gamma_{p,\text{max}}} d\gamma_p \, \gamma_p m_p c^2 \, Q_{p}(N_p, \gamma_p, t)
\]

The spin-down luminosity of the pulsar PSR J0622+3749 is given by $L_{\mathrm{sd}} = 2.7 \times 10^{34}~\mathrm{erg~s^{-1}} \quad$ \cite{Pletsch:2011kp}.The fraction of this luminosity, represented as \(\eta_p = \frac{L_p}{L_{\mathrm{sd}}}\), indicates how much of it is carried by protons to explain the observations. In our case, the model fitting requires that \(\eta_p = 0.14\).

\begin{table}[htb]
\caption{Modeling parameter values of the $pp$ channel to explain the LHAASO J0621+3755 and 3HAWC J0621+382 observations.}
\label{tab:parameters_diffusion}
\centering
%\ruledtabular
\setlength{\extrarowheight}{3pt}
\begin{tabular}{l@{\hspace{1cm}}c}
\hline
Parameters & Values \\
\hline
\hline
$d_L$ (kpc) & 1.6 \\
$t_{\mathrm{age}}$ (kyr) & 207.8 \\
$t_{\mathrm{sd}}$ (kyr) & 10 \\
$D_0$ (cm$^{2}$ s$^{-1}$) & $8.9 \times 10^{27}$ \\
$R$ (pc) & 50 \\
\hline
\hline
$\alpha_p$ & 2.2 \\
$\gamma_{\mathrm{cut}}$ (TeV) & 250 \\
$\gamma_{p,min}$ (TeV) & $1.07\times 10^{2}$\\
$\gamma_{p,max}$ (TeV) & $3.2\times10^{8}$\\
$N_0$ (erg$^{-1}$ s$^{-1}$) & $1.12 \times 10^{39}$ \\
%$t_{\min}$ (sec) & 1 \\
%$E_p$ (TeV) & 160 \\
%$B$ ($\mu$G) & 1\\
\hline
\hline
$L_p$ (erg s$^{-1}$) & $3.76\times 10^{33}$ \\
$\eta_p$ & 0.14\\
\hline
\end{tabular}
\end{table}

\section{\label{Analysis}Summary and Outlook}
In this work, we discussed the sub-PeV $\gamma$-rays observed by LHAASO-KM2A and HAWC from the halo of PSR J0622+3749, focusing on the $pp$ interaction channel. The VHE photons produced through neutral pion decay in this channel typically contribute at energies ranging from a few TeV to over a PeV. 

A previously proposed leptonic model incorporated synchrotron and IC emissions to satisfy the XMM-Newton upper limit and LHAASO-KM2A observations \cite{PhysRevLett.126.241103, Fang:2021qon, VERITAS:2025xjd}. However, this model did not account for the specific emission associated with 3HAWC J0621+382 \cite{HAWC:2020hrt}, assuming that the source was spatially associated with bcu 4FGL J0620.3+3804. To clarify the origin of the TeV emission in 3HAWC J0621+382, we performed a targeted Fermi-LAT analysis of PSR J0622+3749 as a point source, accounting for the pulsar halo within the LHAASO J0621+3755 region and the nearby bcu, 4FGL J0620.3+3804. Our findings indicate that the TeV emission from 3HAWC J0621+382 is not associated with the bcu source, given its energetics. Consequently, when modeling LHAASO J0621+3755 and 3HAWC J0621+382 as a combined pulsar halo, we adopted a hadronic scenario involving neutral pion decay.
Additionally, to complete the lepto-hadronic scenario, we propose that other multiwavelength observations (ranging from X-rays to MeV and a few TeV $\gamma$ rays) can be addressed using the leptonic channel, similar to the modeling presented in \cite{PhysRevLett.126.241103, Fang:2021qon, VERITAS:2025xjd}. However, it is challenging to draw any new conclusions from this channel, as most observations yield upper limits.

This study presents a hadronic model that incorporates a slow diffusion coefficient, $D_0 = 8.9 \times 10^{27}\,\mathrm{cm}^2\,\mathrm{s} ^{-1}$ as proposed in reference \cite{PhysRevLett.126.241103}, along with various other parameters that contribute to the construction of the transport equation for accelerated protons. We solve this transport equation using the GAMERA package. The spectral evolution of the proton population indicates that the majority of protons accumulate in the halo rather than escaping into the interstellar medium, attributed to their extended cooling times relative to their escape times. We then calculated the $\gamma$-ray flux from neutral pion decay using PYTHIA 8.18. In this process, neutral pions are produced through interactions of propagated protons with an ambient medium within a region of 50 pc, with a density of 1 cm$^{-3}$. The proposed $pp$ interaction channel requires an accelerated proton luminosity of 14\% of the spin-down luminosity. Interestingly, the parameter values are consistent with the environmental conditions.

In the HAWC analysis, the source 3HAWC J0621+382 was modeled using a uniform disk morphology with a radius of \(0.5^{\circ}\). In contrast, the flux of LHAASO J0621+3755 was derived from a 2D Gaussian morphology with a standard deviation (\(\sigma\)) of \(0.4^{\circ}\). Therefore, a further flux analysis of 3HAWC J0621+382 using a uniform 2D Gaussian model will be necessary for combined modeling. Regarding the Geminga pulsar halo, both the uniform disk and symmetric 2D Gaussian morphology flux observed by HAWC are available in reference \cite{HAWC:2017kbo} (see figure S2 in that reference). The comparison indicates that the flux is higher for the 2D Gaussian model than for the disk-like morphology. Similarly, a change in morphology may also impact the flux for 3HAWC J0621+382; however, such an analysis is beyond the scope of this paper. It would be interesting to see whether the 2D Gaussian flux of this source from HAWC becomes available in the future, enabling consistent combined modeling.

The strong evidence linking gamma-rays to both leptons and hadrons is derived from observations of neutrinos. The $
$ interaction ultimately produces secondary neutrinos through the decay of charged pions ($(\pi^\pm$)). During this process, the generated neutrinos will have a specific flux. 
$
E_{\nu}^2\frac{dN_{\nu}}{dE_{\nu}}=\frac{2}{3}E_{\gamma}^2\frac{dN_{\gamma}}{dE_{\gamma}}\biggm|_{{E_{\gamma}=2E_{\nu}}}
$. 
Considering the photon differential flux observed by LHAASO, 
\begin{align}
\frac{dN_{\gamma}}{dE_{\gamma}} &= 
(3.11 \pm 0.38_{\text{stat}} \pm 0.22_{\text{sys}}) \times 10^{-16} \nonumber \\
&\hspace{2em} \times \left( \frac{E_{\gamma}}{40 \, \text{TeV}} \right)^{-2.92} 
\, \text{TeV}^{-1} \, \text{cm}^{-2} \, \text{s}^{-1},
\end{align}
 The corresponding neutrino differential flux, without accounting for the parameter errors, can be calculated using:
\begin{equation}
\frac{dN_{\nu}}{dE_{\nu}} \sim 7.5 \times 10^{-18} \left( \frac{E_{\nu}}{100\, \text{TeV}} \right)^{-2.92} \text{TeV}^{-1} \, \text{cm}^{-2} \, \text{s}^{-1} 
\end{equation}

The number of neutrino events expected at IceCube from LHAASO J0621+3755 as a counterpart to the sub-PeV $\gamma$-rays through the pp model \cite{PhysRevD.108.023016} is, 
\begin{equation}
N_{\nu_{\mu}} = \mathbb{T} \int_{\epsilon_{\nu_{\mu}}^{\text{min}}}^{\epsilon_{\nu_{\mu}}^{\text{max}}} \frac{d {N}_{{\nu_{\mu}}}}{d {\epsilon}_{{\nu_{\mu}}}} \, A_{\nu_{\mu}, \text{eff}}(\epsilon_{\nu_{\mu}}) \, d\epsilon_{\nu_{\mu}}
\label{neutrino_number}.
\end{equation}
The period, $\mathbb{T} $, is considered from December 27, 2019, to November 9, 2020, implying a correlated period of  $2.75 \times 10^7$ seconds with the LHAASO observation. The effective area, $A_{\nu_{\mu}, \text{eff}}$ for the full northern sky was used \cite{IceCube:2016ipa} in the above calculation.
The resulting number of neutrino events, $4.54 \times 10^{-3}$ expected in the IceCube Neutrino Observatory for $\epsilon_{\nu_{\mu}}^{\text{min}} = 10^3 \mathrm{GeV}$ and $\epsilon_{\nu_{\mu}}^{\text{max}} = 3.51 \times 10^5 \mathrm{GeV}$. 

Additionally, the neutrino flavor ratio at the source for the $pp$ channel is  $\bar{f}_{\nu_e}^{s}: \bar{f}_{\nu_\mu}^{s}: \bar{f}_{\nu_\tau}^{s} = 1 : 2: 0$ will become $\approx 0.89 : 0.49: 1.62 $,$\approx 1.03:0.57:1.40$ for the neutrino energies, 200 TeV and 1 PeV, respectively, at Earth following oscillation. While a single neutrino source with such detectability may not currently be identified using existing or future detectors, a collection of similar sources can contribute to the Galactic diffuse neutrino flux, as observed recently \cite{IceCube:2023ame}. Future mega-projects like KM3NeT \cite{KM3NeT:2022kzg} and IceCube Gen2 \cite{IceCube-Gen2:2020qha} can further enable significant stacking of diffuse neutrino observations, including their flavor ratios, and support the hadronic channel.
\begin{center}
\textbf{Acknowledgments}
\end{center}
\noindent We are grateful to the anonymous referee for their valuable suggestions, which significantly improved the manuscript.

\bibliography{CITE}

\end{document}